\documentclass[preprint, superscriptaddress, doublespacing]{revtex4-1}
\usepackage{graphicx}
\usepackage{amsmath}
\usepackage{color}
\usepackage{amsfonts}
\usepackage{amssymb}
\usepackage{tabularx}

\newcommand{\kvec}{\boldsymbol{k}}
\newcommand{\qvec}{\boldsymbol{q}}
\newcommand{\lvec}{\boldsymbol{l}}
\newcommand{\Kvec}{\boldsymbol{K}}
\newcommand{\nvec}{\boldsymbol{n}}

\begin{document}
\linespread{1.5}
\begin{titlepage}

\title{Bias Free Gap Creation in Bilayer Graphene}

\author{A. R. Davenport}
\affiliation{Department of Physical Sciences, The Open University, Walton Hall, Milton Keynes MK7 6AA, UK}
%\email{anthony.davenport@open.ac.uk}
\email{Jim.Hague@open.ac.uk}
% Your e-mail will expire over the next few months, so probably better to use mine.

\author{J. P. Hague}
\affiliation{Department of Physical Sciences, The Open University, Walton Hall, Milton Keynes MK7 6AA, UK}

\begin{abstract}
For graphene to be utilized in the digital electronics industry the
challenge is to create bandgaps of order 1eV as simply as
possible. The most successful methods for the creation of gaps in
graphene are (a) confining the electrons in nanoribbons, which is
technically difficult or (b) placing a potential difference across
bilayer graphene, which is limited to gaps of around 300meV for
reasonably sized electric fields. Here we propose that electronic band
gaps can be created without applying an external electric field, by
using the electron-phonon interaction formed when bilayer graphene is
sandwiched between highly polarisable ionic materials. We derive and
solve self-consistent equations, finding that a large gap can be
formed for intermediate electron-phonon coupling. The gap originates
from the amplification of an intrinsic Coulomb interaction due to the
proximity of carbon atoms in neighbouring planes.
\end{abstract}

\maketitle

\end{titlepage}

\section{Introduction}

The experimental discovery of graphene by Novoselov and Geim
\textit{et al.} \cite{Discovery} has led to a major drive to develop
graphene based microelectronics to replace silicon in computing
devices. \cite{Geim2007} Exceptional electrical, transport and
thermodynamic properties make graphene a promising material for this
task, although the honeycomb structure naturally leads to a zero-gap
semiconductor. The goal is to change the electronic structure to make
a useful semiconductor without destroying the properties that make
graphene unique. \cite{Berger2004}

Graphene is both extremely strong and structurally stable, even on
very small length scales, due to $\sigma$ bonds that connect the
carbon atoms together in a hexagonal structure. Therefore, graphene
structures have the potential to function on much smaller nanoscales
than silicon technology.  \cite{Ribbons} Very large graphene sheets
can be manufactured, raising the possibility of creating entire
devices from a single graphene sheet.\cite{Schwierz2010} Energy loss
in such graphene devices would be severely reduced as the number of
connections from one material to another inside the devices are
minimized, which could lead to smaller and faster chips.\cite{Ribbons}

There are many proposals for the creation of gaps in monolayer
graphene, including the alteration of the structural geometry of the
graphene, introduction of impurities (replacing carbon atoms or adding
adatoms), creation of graphene quantum dots \cite{QuantumDots} or by
dimensional reduction by making nanoribbons
\cite{NanoribbonGap}. While these methods are very promising in
theory, current technology and finite size effects put large scale production
outside our reach\cite{Nilsson2008}, but bilayer graphene may offer a
solution.  \cite{BilayerElectricField,CoulombBlocade}

Bilayer graphene, similar to monolayer, is a zero-gap semiconductor,
with valence and conduction bands that meet at the Fermi surface. It
is available in two forms with $AB$ and $AA$
stacking. \cite{Graphene101} If a sufficient electric potential is
applied over the two sheets, a gap opens up at the high symmetry
$\Kvec$ and $\Kvec^{\prime}$ points of the Brillouin zone on the order
of the interlayer hopping energy. \cite{TheoryGap} This effect has
been observed experimentally \cite{ExperimentGap}, where band gaps of
$0.1$-$0.3$eV are seen, but they require large potential differences
to be maintained across the bilayer. Such electronic devices have been
predicted to operate with clock speeds in the order of
terahertz.\cite{TeraHertz} Although the bandgaps described above are
sufficiently large to make digital transistors, the devices still lack
the required properties for large scale production.\cite{Schwierz2010}

Substrates can also modify the electronic properties of epitaxial
graphene \cite{Bostwick2007b} and contribute to changes in the band
structure and resistivity, including electron band gap
creation.\cite{Zhou2007,Enderlein2010} Theoretical works have explored
surface reconstructions for graphene on substrates such as SiC
\cite{Pankratov2010} and possible gap enhancements from interactions
between electrons and surface phonons on polarisable
substrates. \cite{Hague2011}

 The approach proposed in this paper is different to previous schemes
 for opening a gap in bilayer graphene since no potential difference
 across the bilayer is required. While it is often neglected, bilayer
 graphene has a small inherent difference in electrical potential in
 the AB stacking configuration; two of the carbon atoms in the unit
 cell sit directly beneath each other with overlapping electron
 wavefunctions.\cite{Graphene101} We study how this small difference
 is amplified by an electron-phonon interaction between the bilayer
 and a sandwich of a highly polarisable sub/super-strate.

This paper is organised as follows. We introduce a model for unbiased
bilayer graphene including a Holstein electron-phonon interaction in
section \ref{sec:model}, where we also derive self-consistent
equations for the gap. Results showing the evolution of the gap with
electron-phonon coupling strength is presented in section
\ref{sec:results}. We summarise the work in section \ref{sec:summary},
where we also discuss the outlook for observing the gap or its
precursor.

\section{Model and methods}
\label{sec:model}

The electron-phonon interaction has been widely studied in condensed
matter systems, most notably in theories of superconductivity.
\cite{Superconductivity} There have been several studies of its role
in graphene and graphitic structures that are both theoretical
\cite{Cappelluti2012, Stein2013} and experimental \cite{Ulstrup2012,
  Fay2011, Bruna2010}.  This work goes beyond that of previous studies
in an attempt to create a usable and tuneable gap in bilayer graphene
purely by choosing the materials that surround it.

In this paper, {\it we study the electron-phonon interaction between
  electrons in the graphene layer and phonons in a surrounding
  material}. To avoid confusion, we point out that the form of
electron-phonon interaction between electrons in the graphene layer
and phonons that are also in the graphene layer is quite different,
and it is expected theoretically that such interactions vanish at the
K point, essentially due to the symmetries of the interaction and the
phonons (See Refs. \onlinecite{Attaccalite2010a} and
\onlinecite{Park2008a}). ARPES measurements of graphene on a Cu
substrate confirm that as doping decreases, so that the Fermi surface
approaches the K point \cite{Siegel2012}, the dimensionless effective
couping decreases to zero in accordance with the work by Calandra {\it
  et al.}  \cite{Calandra2007}, indicating that the coupling between
graphene and a Cu substrate is very small. On the other hand, the
dimensionless electron-phonon coupling can change dramatically on
different substrates, and for graphene on SiC, the electron-phonon
coupling measured by ARPES \cite{Bostwick2007a} is essentially
independent of the chemical potential as the doping approaches 0.2\%
per site, with an average value for the dimensionless electron-phonon
coupling, $\lambda$, of around $\lambda=0.25$, far larger than the
predictions of in-plane couplings. Since the in-plane interactions
must vanish as doping tends to zero, then any remaining
electron-phonon interaction must be a result of interactions with the
surrounding material. Calandra {\it et al.}  \cite{Calandra2007} find
that the couplings measured by Bostwick {\it et al.} cannot be
explained by an in-plane electron-phonon coupling. In our opinion,
this demonstrates that the substrate is responsible for a completely
different form of electron-phonon coupling between electrons in the
graphene sheet and phonons within the surrounding
material \footnote{We note that it is very difficult to calculate the
  strength of the electron-phonon coupling ab-initio, which can lead
  to large differences between theoretical predictions and
  experimental measurements of $\lambda$ \cite{Yin2013}. Values of
  $\lambda\sim 1$ are surprisingly common in condensed matter systems,
  and have been measured in e.g. Pb and high-$T_{C}$
  superconductors.}.

We model the electronic structure of bilayer graphene by extending the
tight-binding approach that has been highly successful for the study of
free-standing graphene.\cite{Graphene101} A small energy difference,
$\delta$, is induced by the proximity of stacked sites, plus an
additional electron-phonon term representing the effect of introducing
coupling with phonons in the substrate and superstrate. The Hamiltonian
representing this physics is,
\begin{eqnarray}\begin{split}
H = &-\gamma_0 \sum_{\langle \nvec,\nvec^\prime \rangle u \sigma}(X_{\nvec u \sigma}^\dagger Y_{\nvec^\prime u \sigma} + Y_{\nvec^\prime u \sigma}^\dagger X_{\nvec u \sigma}) \\
&-\gamma_1 \sum_{\sigma\nvec} (Y^{\dagger}_{\nvec 1\sigma} Y_{\nvec 2\sigma} + Y^\dagger_{\nvec 2\sigma} Y_{\nvec 1\sigma})  \\
&- g\sum_{\nvec u \sigma} n_{\nvec u \sigma} x_{\nvec u} + \sum_{\nvec} \hbar\Omega(N_{\nvec} + \frac{1}{2}) \label{EQN::Hamiltonian} + \delta\sum_{u, \sigma, \nvec \in Y} n_{\nvec u\sigma}
\end{split}\end{eqnarray}

\begin{figure}[!ht]
\includegraphics[width=0.5\textwidth]{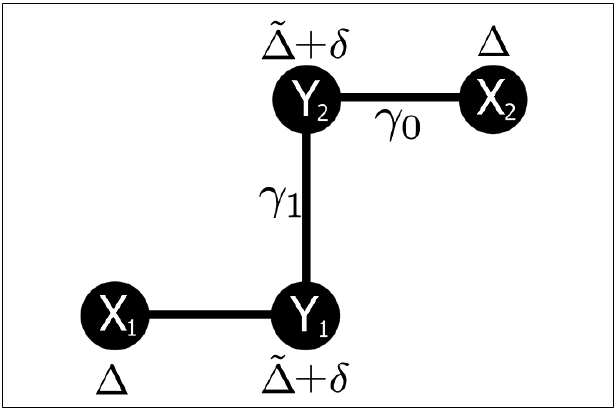}
\caption{A schematic of the physics in the bilayer. Symmetry considerations lead to two different types of site, $X$ and $Y$. There is a hopping $\gamma_0$ between sites in the plane and $\gamma_1$ between Y sites in different planes. Due to the proximity of neighboring planes, the $Y$ sites are increased in energy by a value $\delta$. Real parts of the self energy $\Delta$ and $\tilde{\Delta}$ modify the on-site potential.}
\label{Fig:schematicphysics}
\end{figure}

 Bilayer graphene has four sites in its unit cell, but symmetry in the
 lattice shows that there are only two unique sites, that we call X
 and Y. In eqn. \ref{EQN::Hamiltonian}, $X^{\dagger}_{\nvec u \sigma}$
 and $Y^{\dagger}$ are operators that creates electrons on sites X and
 Y respectively, with spin $\sigma$, on layer $u=\{1,2\}$ on the site
 with lattice vector, $\nvec$ (see
 Fig. \ref{Fig:schematicphysics}). The first and second terms
 represent the contributions of intra-plane and inter-plane hopping to
 the electron kinetic energy. $\gamma_0$ and $\gamma_1$ are the tight
 binding parameters for electrons within graphene planes and between
 graphene sheets respectively. 

The electron-phonon interaction is represented by the third term of
the Hamiltonian. For interactions between electrons confined to
planes, and phonons in surrounding materials, the electron-phonon
interaction must have a form where the local electron density couples
to displacements of ions in the surrounding material. Here, the
standard Holstein approximation is used. $g$ determines
the magnitude of the interaction between the electron occupation
$n_{\nvec u \sigma}$ and an ion in the surrounding material displaced
by a distance $x_{\nvec u}$.

The interaction strength, $g$,
 is directly related to the dimensionless electron-phonon coupling
 $\lambda=g^2/2M\Omega^2\gamma_0$, where $M$ is the ion mass and
 $\Omega$ is the phonon frequency. There is no screening between the
 sub/superstrate and the graphene sheets, which is important for
 obtaining large electron-phonon coupling strengths. The fourth term
 in the Hamiltonian describes phonons in the surrounding material as
 simple harmonic oscillators, where $N_{\nvec}$ is the phonon number
 operator. 

 Strictly, the electron-phonon interaction between the graphene
 bilayer and the substrate (indeed between any electrons confined to a
 layer and surrounding ions) would have a Fr\"ohlich form
 \cite{Alexandrov1999,Alexandrov2002,Steiner2009}, so the results
 presented here are only expected to be qualitatively similar to the
 real system. The local Holstein model is a standard approximation to
 the electron-phonon interaction, chosen here because it significantly
 simplifies the self-consistent equations, while being in the same
 general class of interactions \cite{Hague2011}. When electron-phonon
 coupling is moderate, Holstein and Fr¨ohlich interactions lead to
 qualitatively similar effects on two-dimensional lattices
 \cite{Hague2006,Hague2014}.

Finally we add the crucial small energy difference, $\delta$, between
X and Y sites, which originates from the Coulomb repulsion between Y
sites on different layers. It is these energy differences that seed
the charge density wave state that leads to the band gap.

We compute changes to the effective on-site potentials that occur due
to the introduction of the electron-phonon interaction.  Perturbation
theory was used to obtain self-consistent equations for the change in
on-site potentials, a technique that can be highly accurate at low
phonon frequency. The order two rotational symmetry means that the
simplest solution requires two modified on-site potentials, $\Delta_n$
and $\tilde{\Delta}_n$. The two functions represent interaction in
sites of type X, that only have hopping terms to other atoms in their
own layer ($\Delta_n$) and sites of type Y, that interact with the
other layer ($\tilde{\Delta}_n$). The energies involved in the bilayer
are summarized in Fig. \ref{Fig:schematicphysics}.

We derive self-consistent equations using a general form of the self-energy, which is consistent with the symmetry of the problem:

\begin{widetext}\begin{gather}
\Sigma(i\omega_n) \approx 
\left( \begin{array}{cccc}
i\omega_n(1-Z_n)+\Delta_n & 0 & 0 & 0 \\
0 & i\omega_n(1-\tilde{Z}_n)+\tilde{\Delta}_n & 0 & 0 \\
0 & 0 & i\omega_n(1-\tilde{Z}_n)+\tilde{\Delta}_n & 0 \\
0 & 0 & 0 & i\omega_n(1-Z_n)+\Delta_n
\end{array} \right)
\end{gather}\end{widetext}

Here $Z_n$ and $\tilde{Z}_n$ are the quasi-particle weights, that
represent change in the effective mass for the gap functions
$\Delta_n$ and $\tilde{\Delta}_n$ respectively. Both $\Delta_n$ and
$Z_n$ are functions of the fermion Matsubara frequency $\omega_n =
2\pi k_BT(n+1/2)$, where $n$ is an integer and $T$ is the
temperature. Off diagonal terms in the self-energy are zero in the
perturbation theory considered here and terms are taken to be momentum
independent. This is consistent with a local approximation or
dynamical mean-field theory (DMFT). 
%In such cases, the interaction is
%also momentum averaged so the approximation used here is also
%consistent with a Fr\"{o}hlich interaction with the same value of
%$\lambda$ (momentum is not conserved at internal vertices in DMFT /
%local approximation, so Holstein and Fr\"ohlich interactions are
%identical \cite{Muller-Hartmann1989}).

We construct the full Green's function of the system using Dyson's
equation,
$\mathcal{G}^{-1}(\kvec,i\omega_n)=\mathcal{G}_0^{-1}(\kvec,i\omega_n)-\Sigma(i\omega_n)$,
where the non-interacting Green's function of the system can be found
from $\mathcal{G}_0^{-1}(k,i\omega_n)=[Ii\omega_n-H]$:

\begin{gather}
\mathcal{G}^{-1}_0(\kvec,i\omega_n) = 
\left( \begin{array}{cccc}
i\omega_n & \Phi_{\kvec} & 0 & 0 \\
\Phi^*_{\kvec} & i\omega_n-\delta & \gamma_1 & 0 \\
0 & \gamma_1 & i\omega_n-\delta & \Phi_{\kvec} \\
0 & 0 & \Phi^*_{\kvec} & i\omega_n
\end{array} \right)
\end{gather}
$\Phi=\gamma_0\sum_{\lvec}e^{-i\kvec \cdot \lvec}$ 
describes the electron in plane hopping, where we sum over the
nearest neighbor vectors, $\lvec$.\cite{Graphene101} 
Self consistent equations for the effective on-site potentials are derived by inverting the Green's function
matrix and then placing it into the 1st order contribution to the
self-energy

\begin{eqnarray}
\Sigma_{ij}(\kvec,i\omega_n) = -\gamma_0\lambda  k_BT\sum_{i\omega_s}\int \frac{d^2\qvec}{V_{BZ}}\mathcal{G}_{ij}(\kvec-\qvec,i\omega_{n-s})d_0(\qvec,\omega_s),
\label{EqnLowOrderPert}
\end{eqnarray}
where the phonon propagator $d_0(i\omega_s)=\delta_{ij}\Omega^2/(\Omega^2-\omega_s^2)$, $\omega_s=2\pi k_B T s$ and $s$ is an integer.

We obtain two simultaneous equations from Equation
(\ref{EqnLowOrderPert}) that describe how the effective potential
changes with our free parameters; temperature, phonon
frequency, on-site potential and electron-phonon coupling constant.  

\small
\begin{widetext}\begin{gather}
\Sigma_{X}(i\omega_n) = \gamma_0\lambda k_BT\sum_f [2d_0(i\omega_{s=0})-d_0(i\omega_s)] \int d\varepsilon D(\varepsilon) \frac{\tilde{\Pi}_f \varepsilon^2 - \Pi_f (\tilde{\Pi}_f^2 - \gamma_1^2)}{(\varepsilon^2 + \Pi_f( \gamma_1-\tilde{\Pi}_f))(\varepsilon^2 + \Pi_f(\gamma_1+\tilde{\Pi}_f))} \label{Eqn:Delta} \\ 
\Sigma_{Y}(i\omega_n)  = \gamma_0\lambda k_BT\sum_f [2d_0(i\omega_{s=0})-d_0(i\omega_s)] \int d\varepsilon D(\varepsilon) \frac{\Pi_f \varepsilon^2 - \tilde{\Pi}_f\Pi_f^2}{(\varepsilon^2 + \Pi_f( \gamma_1-\tilde{\Pi}_f))(\varepsilon^2 + \Pi_f(\gamma_1+\tilde{\Pi}_f))} \label{Eqn:DeltaBar}
\end{gather}\end{widetext}\normalsize

$D(\varepsilon)$ is the monolayer density of states (DOS) and
$f=n-s$. The equations were simplified by the substitution of $\Pi_f =
\Delta_f - i\omega_fZ_f$, $\tilde{\Pi}_f = \tilde{\Delta}_f - \delta -
i\omega_f\tilde{Z}_f$, $\Sigma_{X}(i\omega_n) = \Delta_n + i \omega_n
(1-Z_n)$ and $\Sigma_{Y}(i\omega_n) = \tilde{\Delta}_n + i \omega_n
(1-\tilde{Z}_n)$. $\Delta_f$,$\tilde{\Delta}_f$,$Z_f$ and
$\tilde{Z}_f$ are all taken to be real so each equation is solvable by
separation of real and imaginary parts.

%Summing over all Matsubara frequencies, truncated at sufficiently large $\omega_n$ to ensure convergence ($\omega_{n_{max}}=190\gamma_0$), the equations can be solved self-consistently. 

\section{Results}
\label{sec:results}

Equations (\ref{Eqn:Delta}) and (\ref{Eqn:DeltaBar}) were solved
numerically with a linear approximation to the monolayer DOS, to
reduce computation cost. Such an approximation is valid for energies
close to the Fermi energy, and differs from the full DOS by less than
$1\%$.

Tight binding parameters of $\gamma_0=3$eV, $\gamma_1=0.1\gamma_0$ and
$\delta=0.007\gamma_0$ were used to mimic experimental
values.\cite{ParameterValues} Temperatures of $k_BT = 0.01\gamma_0$
and $k_BT = 0.02\gamma_0$ were explored corresponding to $324K$ and
$648K$ respectively. Phonon frequencies of $\Omega = 0.01\gamma_0 =
30$meV and $\Omega = 0.02\gamma_0 = 60$meV were also investigated.

\begin{figure}[!ht]
\centering
\begin{tabular}{cc}
\includegraphics[width=0.35\textwidth]{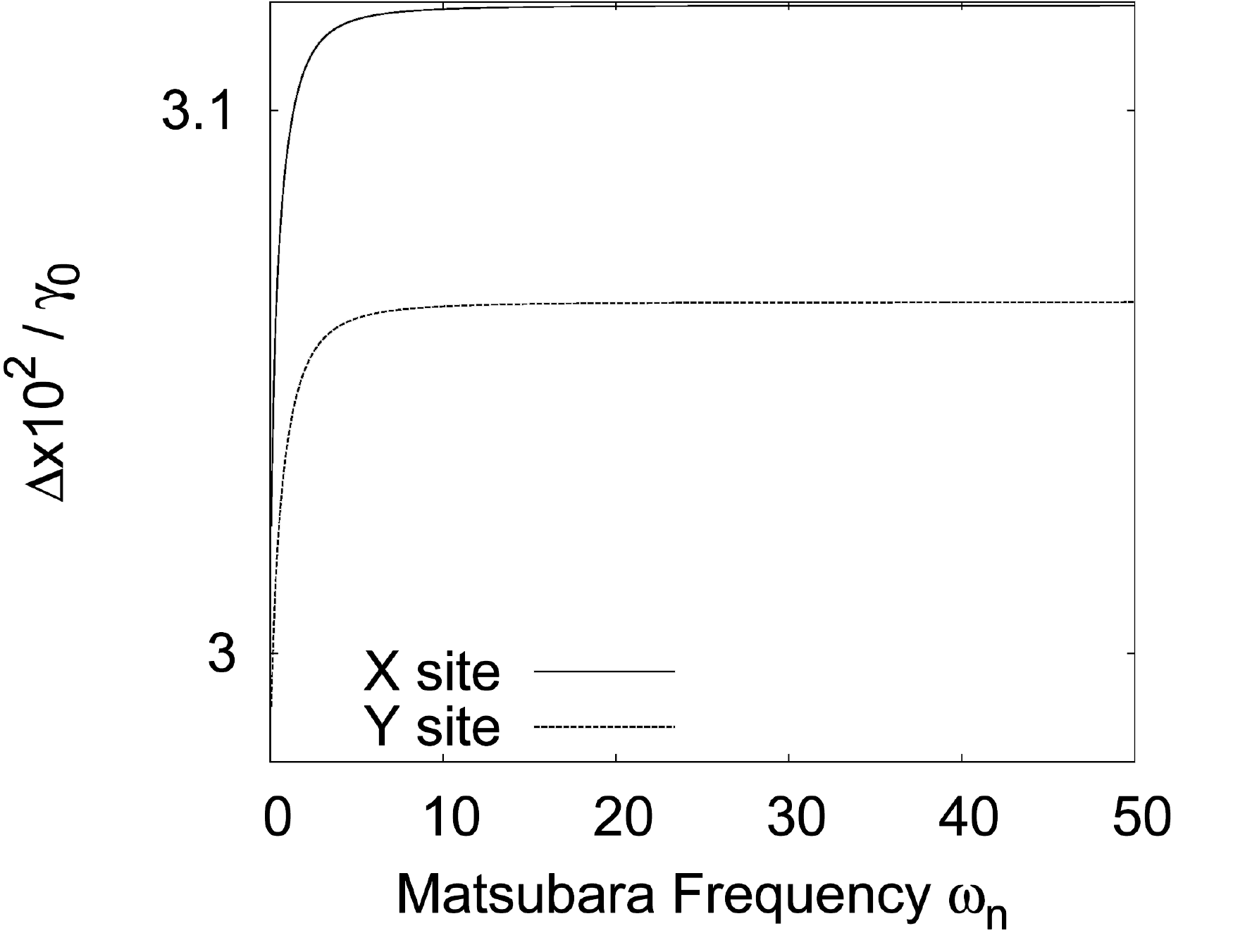} & 
\includegraphics[width=0.35\textwidth]{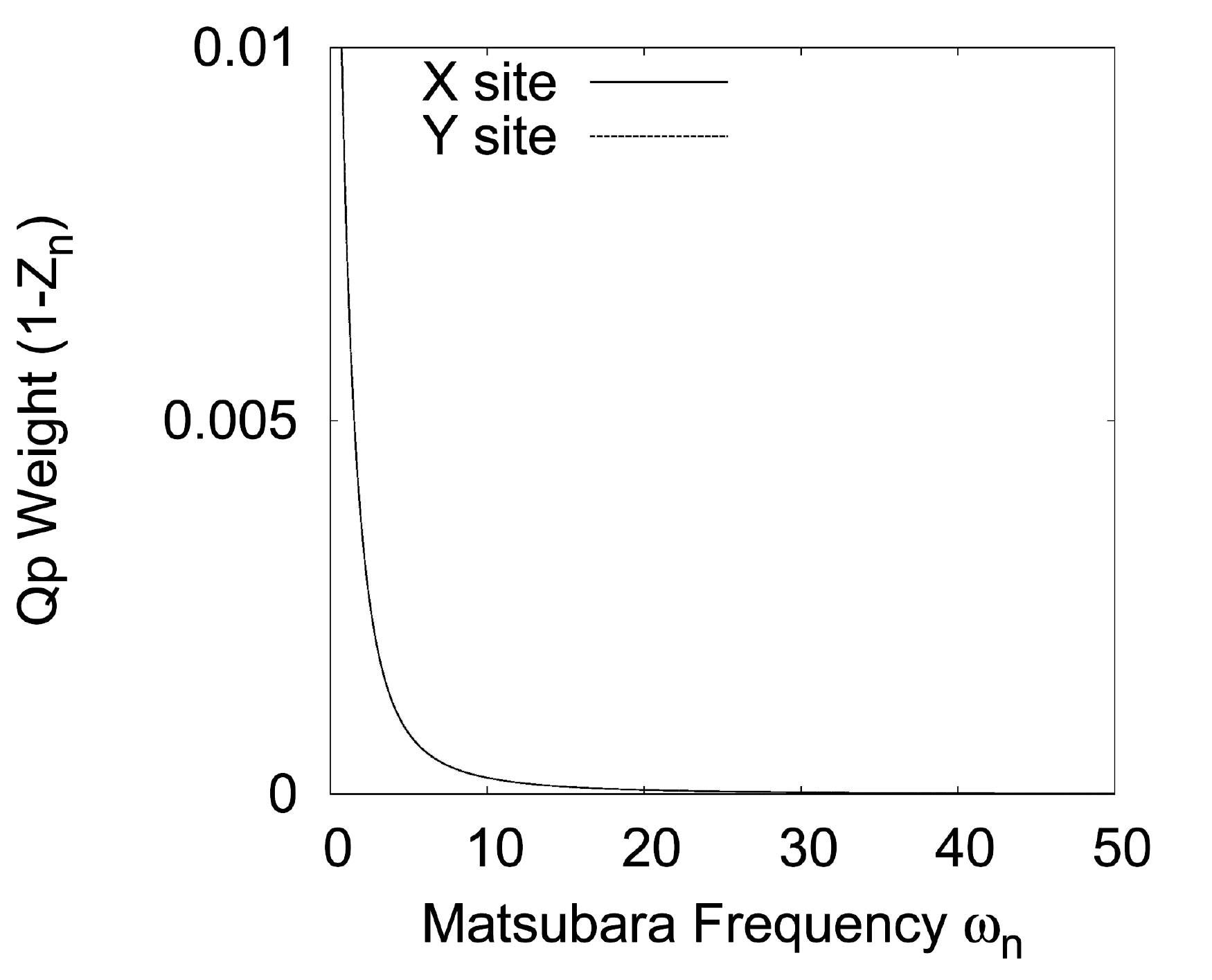} \\
\includegraphics[width=0.35\textwidth]{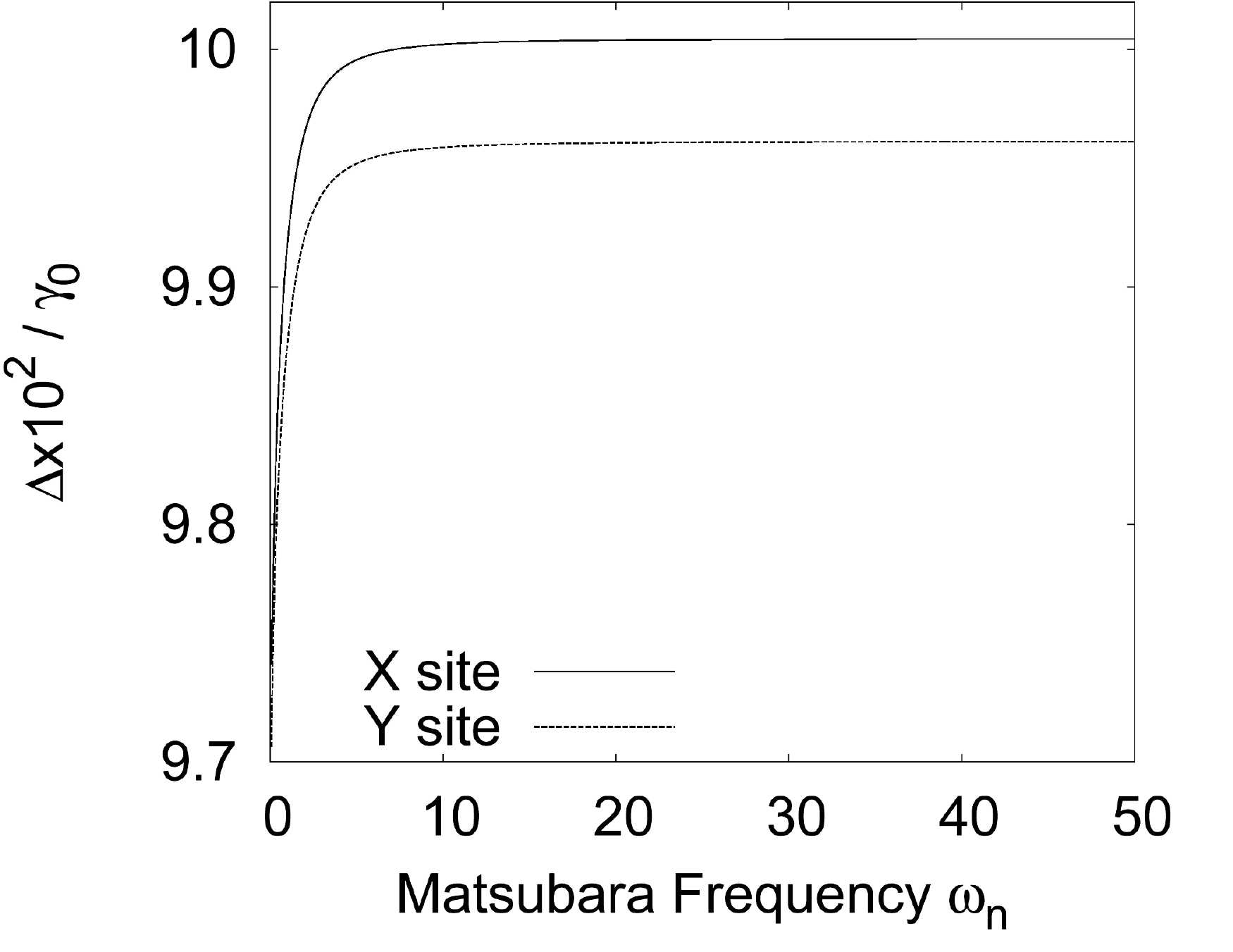} & 
\includegraphics[width=0.35\textwidth]{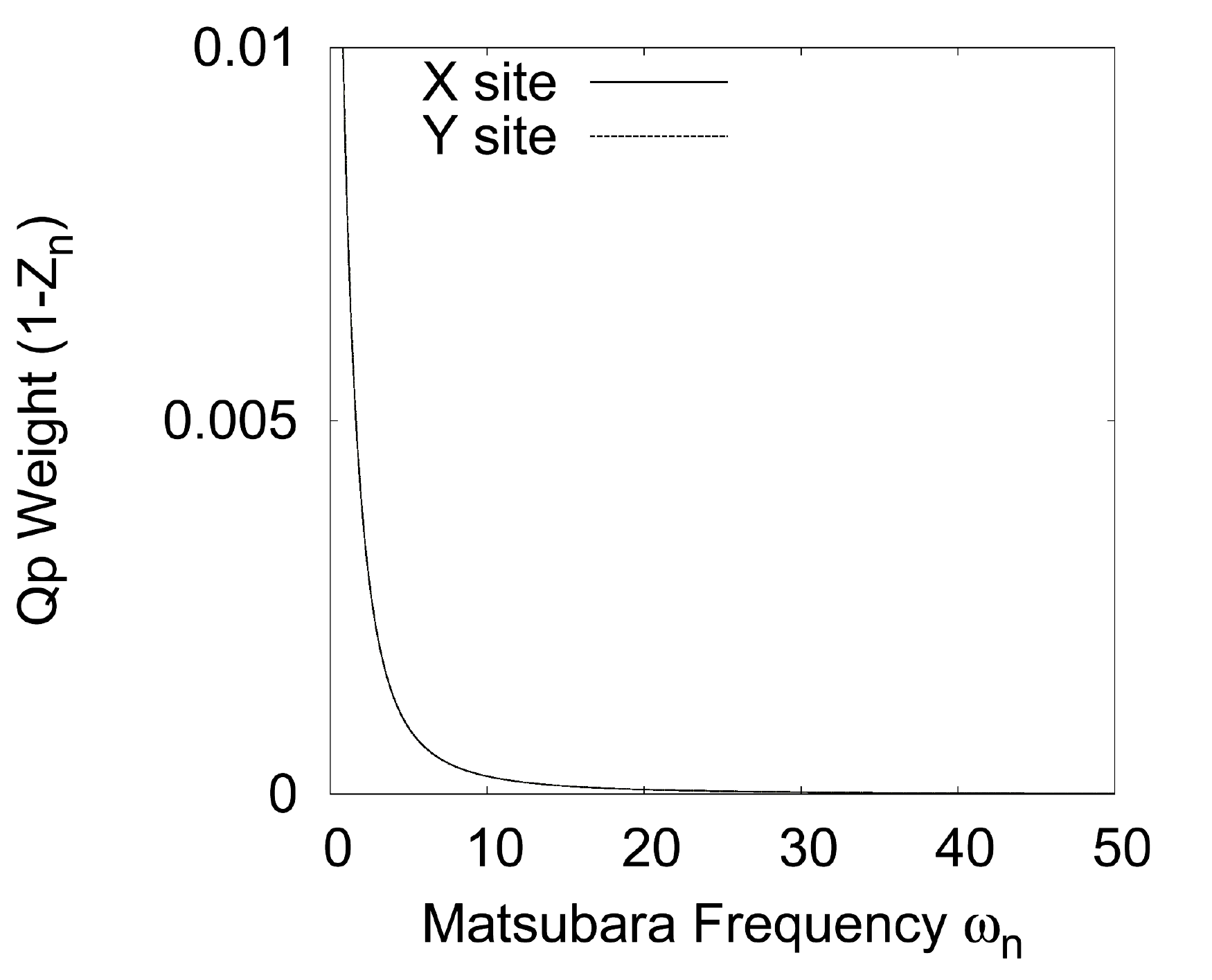} \\
\end{tabular}
\label{Fig:Matsubara}\caption{Matsubara frequency dependance of the real part of the self-energy (on-site correction to the local potential, $\Delta(i\omega_n)$) and its associated quasi-particle weight $Z(i\omega_n)$. Values of electron phonon coupling are $\lambda = 1.1$ (top panels) and $\lambda = 1.2$ (bottom panels), representative of the values of coupling corresponding to the gap opening. Both functions are only weakly dependent on the Matsubara frequency, with the on-site correction to the local potential varying by around 2\% and the quasi-particle weight varying by around 1\%.}
\end{figure}

The Matsubara frequency dependance of solutions to Equations
(\ref{Eqn:Delta}) and (\ref{Eqn:DeltaBar}) is shown in Figure
\ref{Fig:Matsubara}. The electron-phonon coupling strengths used here
are indicative of the interactions investigated in this paper:
$\lambda = 1.1$ (Top) and $\lambda = 1.2$ (Bottom). In both cases the
real part of the self energy for both X and Y sites (which represents
the correction to the on-site potential energy), $\Delta(i\omega_n)$,
increases rapidly before levelling off to a constant value by
$\omega_n=50$. The associated quasi-particle weight $Z_n(i\omega_n)$
tends to a value of one as the potential function approaches
convergence. Sums in the self consistent equations were cut-off at
around $\omega_n=380$ to ensure numerical convergence over all
electron-phonon coupling constants. Both functions are only weakly
dependent on the Matsubara frequency, and vary by less than 2\% over
the range of frequencies. The flat tail at large Matsubara frequency
means that the frequency independent parts of the local potentials can
be determined for calculation of the real frequency band structure.

\begin{figure}[!ht]
\includegraphics[width=0.7\textwidth]{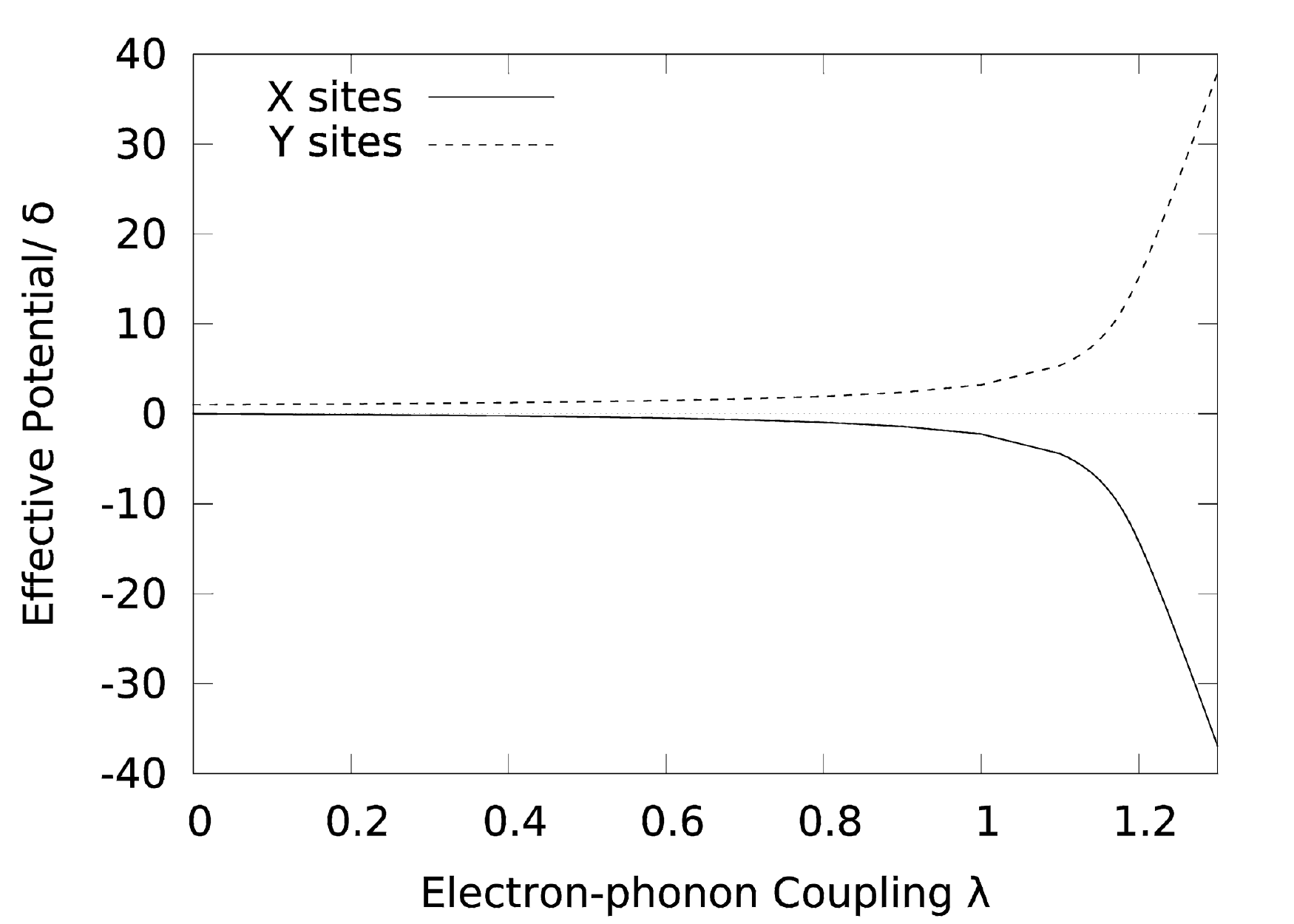}
\caption{Bilayer graphene effective potential normalized to $\delta$,
for Y sites (with intra-layer hopping) and X sites (with both intra and
interlayer hopping). As the electron-phonon coupling is increased, a
near equal and opposite potential forms on Y sites compared to X sites,
 with values of $\Delta = 15\delta$ at $\lambda = 1.2$.}
\label{Fig:EffectivePotential}
\end{figure}

The evolution of the on-site potentials $\Delta/\delta$ (note that
$\Delta = \Delta_{n\rightarrow\infty}$) with increasing
electron-phonon coupling is shown in
Fig. \ref{Fig:EffectivePotential}. It can be seen that the addition of the
electron-phonon coupling leads to an increase in the effective
$\Delta$ for both X and Y sites.  This occurs as follows: the small
value $\delta$ on the Y sites is amplified by the electron-phonon
coupling. An equivalent but opposite effective potential is formed on
X sites, which is the same except for the small difference $\delta$,
since charge is conserved during self-consistency.  The effective
on-site potential has increased to $\Delta\approx 3\delta$ by $\lambda
= 1$, and then goes through a rapid increase around $\lambda = 1.3$
with enhancements exceeding $\Delta=35\delta$. The gap enhancement can
be interpreted as an instability to charge density wave order, where
electrons are more likely to be found on X sites. The small potential
difference between the two sites brings about a variation in the
electron density that is amplified by the electron-phonon
interaction. Although not fixed in the self-consistency, examination
of the figure indicates that the potentials $\Delta_n$ and
$\tilde{\Delta}_n$ are related via $\Delta_n=\delta-\tilde{\Delta}_n$.

\begin{figure}[!ht]
\centering
\includegraphics[width=0.55\textwidth]{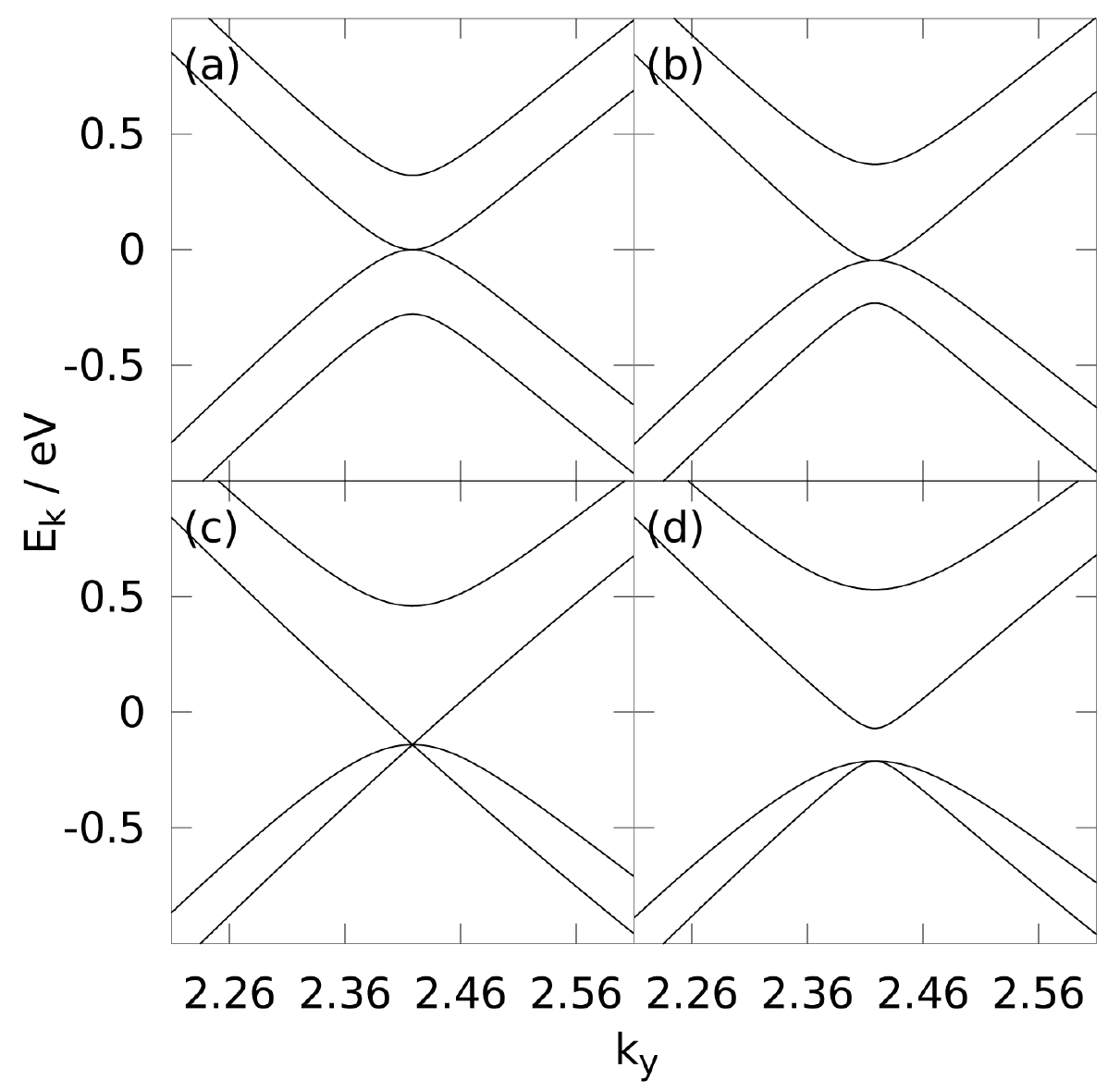}
\includegraphics[width=0.5\textwidth]{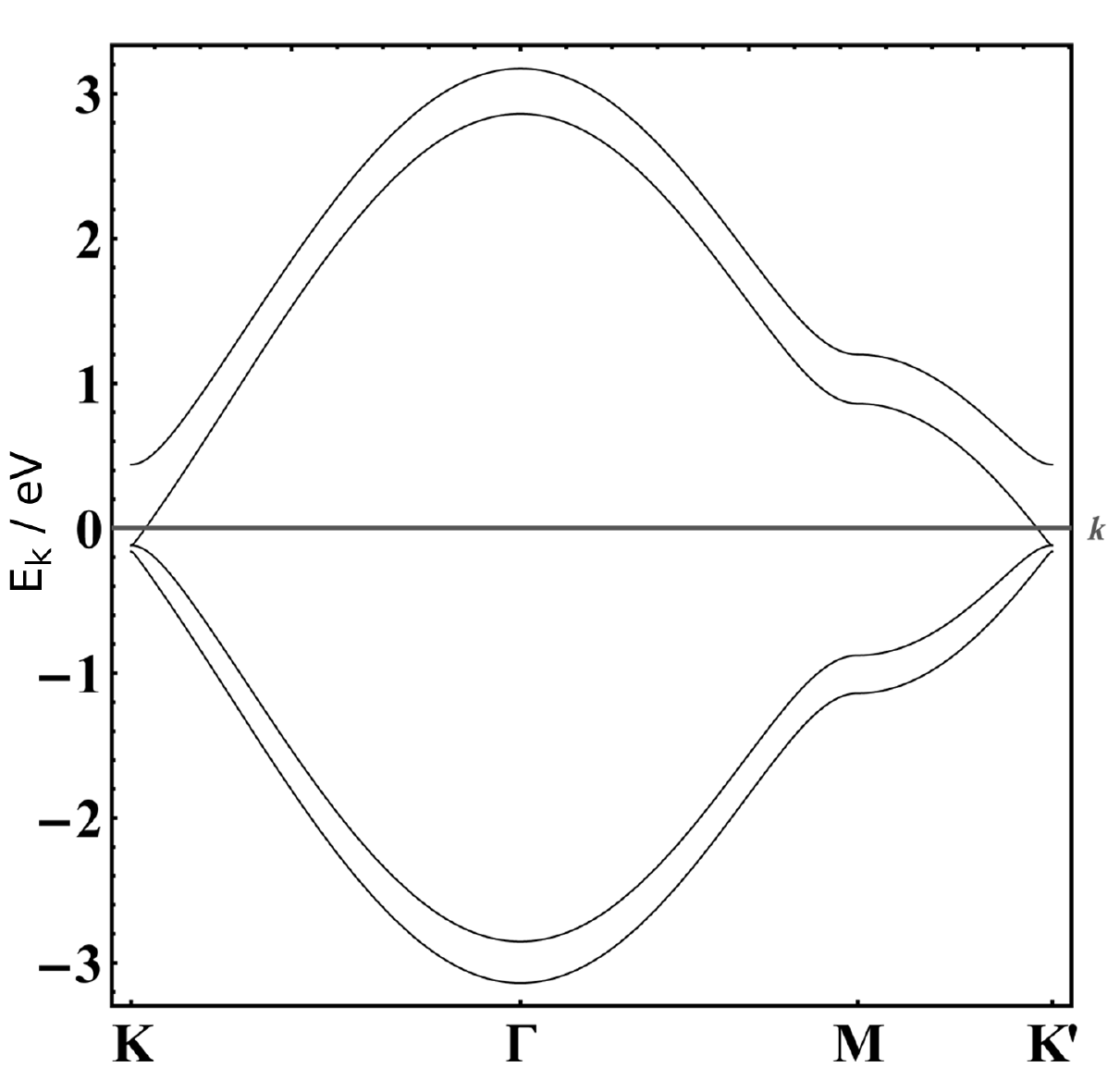}
\caption{(Upper panel) Band structure of unbiased bilayer graphene near the K point showing band gap creation with increasing $\lambda$. The Band structure is displayed at, (a) $\lambda=0$ for bare bilayer graphene ($\Delta(i\omega_n)=0$), (b) $\lambda=1.0$  ($\Delta=3\delta$), (c) $\lambda=\lambda_{\mathrm{crit}}=1.14$ ($\Delta=\Delta_{\mathrm{crit}}$) just before the gap is created and (d) $\lambda=1.17$ ($\Delta=10\delta$) shortly after gap creation. (Lower panel) Band structure along the high symmetry directions at $\lambda_{\mathrm{crit}}=1.14$.}
\label{Fig:ElectronBandStructure}
\end{figure}

To demonstrate how this leads to band gap creation,
Fig. \ref{Fig:ElectronBandStructure} displays the electron band
structure close to the K point, for various values of the
electron-phonon interaction strength. The bands are calculated along
the line $k_x=0$ (which is also a point at which Dirac cones are
observed in monolayer graphene), and is representative of all the
high-symmetry points at the Fermi energy. We note that due to symmetry
considerations, the effective Hamiltonian for low energy excitations
of any system with the bilayer structure will have the same form, even if
the origin of the terms in the Hamiltonian are different (for example,
the same low energy features due to static Coulomb potentials formed
between a graphene monolayer, buffer carbon layer and a SiC substrate
can be seen close to the Fermi surface in Ref. \cite{Pankratov2010})
but we show the low energy band structure here to assist the reader to
identify the features that may be seen experimentally in the context
of specific values of electron-phonon coupling. Figure
\ref{Fig:ElectronBandStructure}(a) demonstrates the bare bilayer
graphene band structure when the electron-phonon coupling is zero and
no gap is present, so only $\delta$ is included. Increase in the
electron-phonon coupling to $\lambda=1.0$ (panel b) shifts the lower
bands down in energy, but does not lead to qualitative change in the
band structure.The lowest energy bands drop slowly with increased
$\lambda>\lambda_{\mathrm{crit}}$ until the two bottom bands touch at
$-\gamma_1/2$, and Dirac cones are created. Figure
\ref{Fig:ElectronBandStructure}(c) is the bilayer graphene band
structure when $\lambda=\lambda_{\mathrm{crit}}=1.14$ where the system
is on the verge of opening a gap, which occurs when the on-site
potentials have the value $\Delta_{\mathrm{crit}} =
(\gamma_1-\delta)/2$, and dominate the band structure. For $\Delta >
\Delta_{\mathrm{crit}}$ the band gap opens up monotonically with
increasing $\Delta$. Figure \ref{Fig:ElectronBandStructure}(d) shows
the gap that has opened when $\Delta = 10\delta$. Strictly, the small
Matsubara frequency dependence of $\Delta$ and $\tilde{\Delta}$ means
that the energies in the band structure will be slightly broadened so
that the quasi-particles have a finite lifetime, so the band
structures shown here are a (close) approximation to the true spectrum
of the Green's functions calculated self-consistently. The lower panel
shows the band structure along the high symmetry directions in the
Brillouin zone. The main changes to the band structure due to the
electron-phonon interaction studied in this paper occur close to the
$K$ point.

\begin{figure}[!ht]
\includegraphics[width=0.7\textwidth]{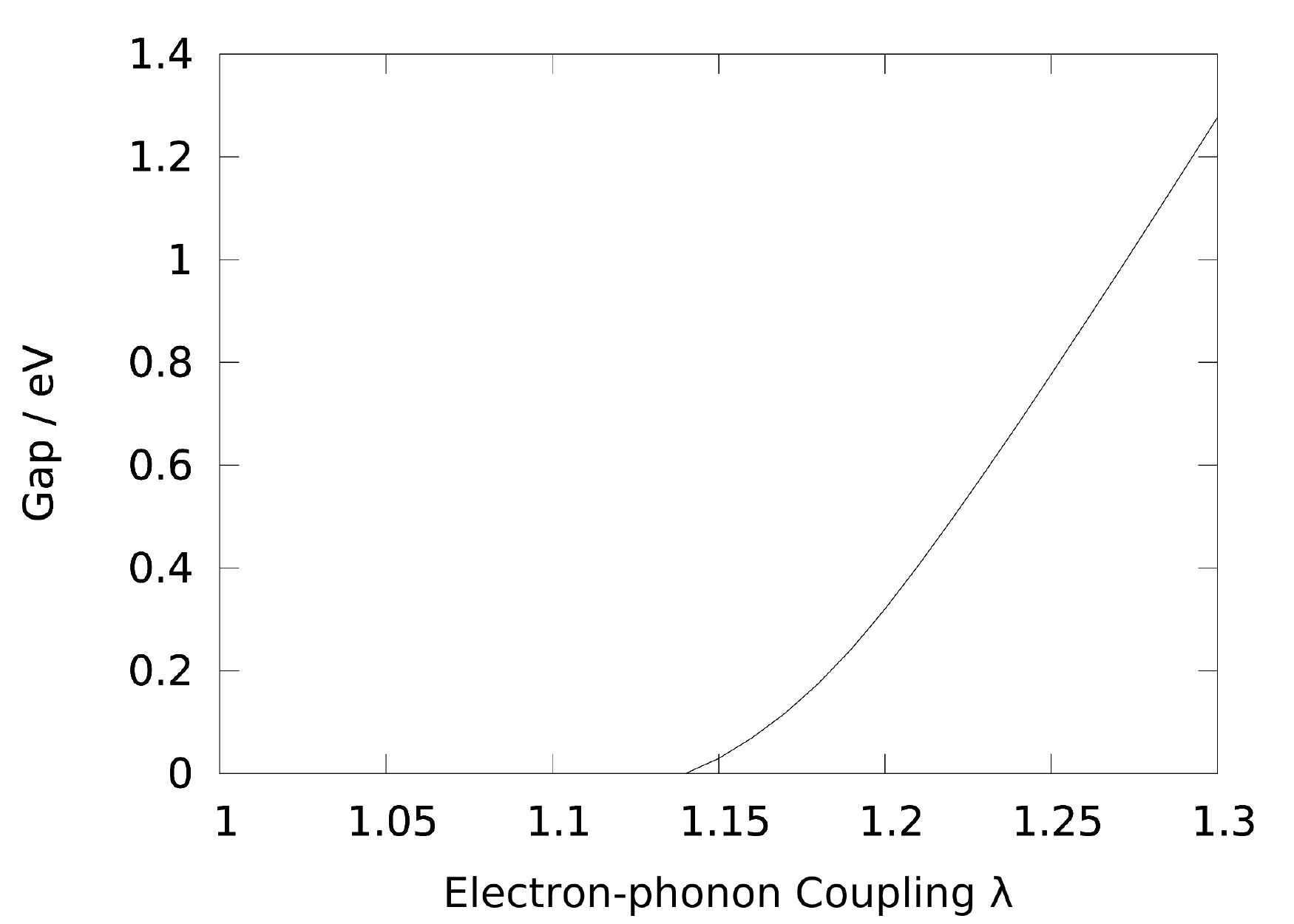}
\caption{Formation of a bias-free gap in bilayer graphene. The gap
opens at $\lambda \approx 1.14$, and increases rapidly with
electron-phonon coupling. A gap of $E_g \approx 1eV$ is found for
dimensionless electron-phonon coupling constants of around
$\lambda=1.25.$}
\label{Fig:EffectivePotentialBias}
\end{figure}

Figure \ref{Fig:EffectivePotentialBias} plots the resulting band gap
created in the range $\lambda=1.1-1.3$, as this is where a gap is
first formed. Figure \ref{Fig:EffectivePotentialBias} shows that the
electron-phonon interaction breaks the zero gap semiconductor
structure of bilayer graphene. A gap is formed at $\lambda \approx
1.14$, that increases rapidly and exceeding that of biased bilayer
graphene (which has a plateau when $E_g \approx 300$meV) at around
$\lambda = 1.2$. Referring back to the definition of
$\Delta_{\mathrm{crit}}$, we note that an increase in $\delta$ or a
decrease in $\gamma_1$ would result in an lower electron-phonon gap
transition.

The local approximation to the interactions used here is also
approximately valid for longer range Fr\"ohlich interactions. Detailed
dynamical cluster approximation calculations on monolayers indicate
that while enhancement effects are reduced by long range interactions,
they are increased by introducing the phonon self-energy (which is not
included here to keep the self-consistent equations simple)
\cite{Hague2014}, such that the overall enhancement predicted here is
expected to be broadly accurate.  We note that a pure Holstein
interaction might be introduced by functionalizing the bilayer with
appropriate molecules, since the bonds from the functionalization
would act as local oscillators. In such systems, 50\% coverage that
differentiates between A and B sites could lead to a larger $\delta$,
thus making it easier to open the gap.

\section{Summary}
\label{sec:summary}

In summary, we have investigated the modification of the band
structure of unbiased bilayer graphene when electron-phonon coupling
is introduced by sandwiching the bilayer graphene with a highly
polarizable material. A perturbative analysis was used to calculate
the development of a band gap as the electron-phonon coupling constant
is varied. We found
that with sufficiently large coupling constants, an electron band gap
could be created in bilayer graphene without the need for an external
potential. There is a large range of band gaps up to $E_g =1.3$eV
within the range $\lambda=1.1-1.3$. These gaps surpass those
achievable in biased bilayer systems. It is interesting to note that
the addition of a potential difference across the bilayer system
breaks the symmetry of the bilayer, and would potentially destroy the
gap creating an effective on/off switch in a similar manner to that of
a field effect transistor.

Obtaining large values of $\lambda$ would be experimentally
challenging.  It has previously been demonstrated that graphene can be
transferred to arbitrary substrates \cite{Caldwel2010} so it is
possible that a suitable value of $\lambda$ could be found optimal
choice of the material surrounding the bilayer. In our opinion, this
would be achievable since a lot of the experimental effort to date has
focused on reducing the interactions with substrates as much as
possible to observe the properties of free-standing graphene (such as
high mobilities), rather than searching for the highest possible
coupling. In spite of this, intermediate electron-phonon coupling
strengths between graphene and neighboring materials have been
measured for various substrates (including SiC) with electron-phonon
coupling that is essentially filling independent at weak dopings of
around 0.2\%, with an average value of $\lambda\approx0.25$
\cite{Bostwick2007a,Gruneis2009}. With a slight increase in doping,
experiments indicate that the electron-phonon coupling ranges from 0.2-0.6 and could even be
as high as 1 \cite{Siegel2012}, so the values required to open a gap
are only slightly bigger than the largest that have been
experimentally observed (since $\lambda\propto g^2$, a small increase
in $g$ would suffice). In addition, the value of $\lambda$ might be
increased by placing the proposed sandwiched sheets under pressure to
reduce the distance between graphene and substrate (the pressure would
also increase $\delta$ and the other tight binding parameters to
varying degrees). Initially, proof-of-concept experiments should focus
on the opening of gaps when there is very large coupling to a highly
polarizable ionic material surrounding the bilayer, before refining
this gap with the use of e.g. spacer layers. Clearly, such systems
warrant further study to determine their full potential.

\section*{Acknowledgments}
We would like to thank Z.Ma\v{s}\'{i}n, N.Mason, A.Umerski and J.Fine for useful discussions. JPH would like to acknowledge EPSRC grant No. EP/H015655/1.

\bibliographystyle{unsrt}
\bibliography{BilayerGraphene}

\end{document}